\documentclass{article}

\usepackage{dsfont,amscd,amsmath,amssymb}
\def\DH{\rm I\kern-1.5pt\rm H\kern-1.5pt\rm I}

\newcommand{\bQ}{{\overline Q}{}}
\newcommand{\bS}{{\overline S}{}}
\newcommand{\bZ}{{\overline Z}{}}

\def\DR{\rm I\kern-1.45pt\rm R}
\def\DC{\kern2pt {\hbox{\sqi I}}\kern-4.2pt\rm C}

\newcommand{\mq}{\mathbf{q}}
\newcommand{\mpsi}{\boldsymbol{\psi}}
\newcommand{\mlambda}{\boldsymbol{\lambda}}

\newcommand{\cD}{{\cal D}}

\newcommand{\bpsi}{{\bar\psi}}

\newcommand{\nn}{\nonumber}
\newcommand{\ba}{\begin{array}}
\newcommand{\ea}{\end{array}}
\newcommand{\be}{\begin{equation}}
\newcommand{\ee}{\end{equation}}
\newcommand{\bea}{\begin{eqnarray}}
\newcommand{\eea}{\end{eqnarray}}
\newcommand{\bi}{\begin{itemize}}
\newcommand{\ei}{\end{itemize}}

\newcommand{\eps}{\varepsilon}

\newcommand{\p}[1]{(\ref{#1})}

\begin{document}\thispagestyle{empty}
\thispagestyle{empty}
\vspace{2cm}
\begin{flushright}
\end{flushright}\vspace{2cm}

\begin{center}
{\Large\bf Partial breaking of global supersymmetry and super particle actions}
\end{center}
\vspace{1cm}

\begin{center}
{\large\bf S.~Bellucci${}^a$, N.~Kozyrev${}^b$, S.~Krivonos${}^{b}$,
A.~Sutulin${}^{a,b}$ }
\end{center}

\begin{center}
${}^a$ {\it
INFN-Laboratori Nazionali di Frascati,
Via E. Fermi 40, 00044 Frascati, Italy} \vspace{0.2cm}

${}^b$ {\it
Bogoliubov  Laboratory of Theoretical Physics, JINR,
141980 Dubna, Russia} \vspace{0.2cm}

\end{center}
\vspace{2cm}

\begin{abstract}\noindent
We argue the conjecture that the on-shell component  super particle actions have a universal form, in which the physical fermions enter the action  
through the ein-bein  and the space-time derivatives of the matter fields, only. We explicitly constructed the actions for the super particles in 
$D=3$ realizing the $N=4\cdot 2^{k} \rightarrow N=2\cdot 2^k$ pattern of supersymmetry breaking, and in $D=5$ with the $N=16$ 
supersymmetry broken down to the $N=8$ one.
All constructed actions have indeed a universal form, confirming our conjecture.
Our construction is strictly based on the assumption that in the system we have one half breaking of the global
supersymmetry, and on the very special choice of the superspace coordinates and component fields.
\end{abstract}

\newpage
\setcounter{page}{1}
\setcounter{equation}{0}
\section{Introduction}
It has been advocated in \cite{BKS1} that the component on-shell actions for the theories with one half spontaneous
breaking of global supersymmetries have an extremely simple form, being written in terms of proper physical components.
The limitation to the theories with one half breaking of supersymmetries is very important, because just in these cases
all physical fermions are Goldstone fermions, accompanying the supersymmetry breaking. If the algebra of the
extended supersymmetry has the following form\footnote{For the sake of brevity we suppress here all space-time and
internal symmetries indices.}
\be\label{1}
\left\{ Q,Q\right\} \sim P, \; \left\{ S,S\right\} \sim P, \; \left\{ Q,S\right\} \sim Z,
\ee
where the $Q$ supersymmetry is supposed to be unbroken, while $S$ supersymmetry and the central charges $Z$ symmetry will be treated 
as spontaneously broken ones, then one may realize all these symmetries by the left action on the coset element
\be\label{2}
g=e^{x P} e^{\theta Q} e^{\mpsi S} e^{\mq Z}.
\ee
Here, $\mq(x,\theta)$ and $\mpsi(x,\theta)$ are Goldstone superfields. Due to the fact that $\# Q=\# S$, {\it all
physical fermions } in the system are just the first components of the superfield $\mpsi:\;\psi=\mpsi |_{\theta=0}$.

The choice of the coset element \p{2} is very important, because the variation of $\theta$ under the spontaneously
broken $S$ supersymmetry is zero, while the superfield $\mq$ transforms as follows
\be\label{3}
\delta_S \theta =0, \qquad \delta_S \mq \sim \eta \theta.
\ee
Therefore, if we are concentrating on the $S$ supersymmetry, then we may replace the coset \p{2} by its $\theta=0$ part
\be\label{2a}
g_{\theta =0} =e^{x P}  e^{\psi S} e^{q Z}, \qquad q=\mq |_{\theta=0}.
\ee
Clearly, the transformation properties of the fields $q$ and $\psi$ under $S$ supersymmetry will be generated
by the left action of the element $g_0=\exp(\eta S)$ on this coset. Now, it  follows from the commutation relations \p{1} that 
$\psi$ transforms under $S$ as the Goldstino in the Volkov-Akulov model \cite{VA}, while
$q$ may be treated as the matter field $(\delta_S q=0)$. Therefore, the physical fermionic components may enter the component 
on-shell action through the d-bein ${\cal E}$, constructed with the help of the Cartan form $\Omega=g^{-1}_{\theta =0}dg_{\theta =0}$, 
and through the space-time derivatives of the matter fields ${\cal D}_x q$, only.

The above considerations strictly fix the possible form of the component on-shell actions. Of course, when we are
dealing with the system containing the gauge fields (which cannot be associated with the parameters of the proper coset) the 
situation becomes more complicated. But if we instead limit ourselves to the one-dimensional case, where we
have only one coordinate $t$, the situation is greatly simplified because
\begin{itemize}
\item there are no gauge fields in $d=1$;
\item all scalar fields are coordinates for the central charges in the superalgebra \p{1};
\item the einbein ${\cal E}$ and the covariant derivatives ${\cal D}_t$ do not carry any indices.
\end{itemize}
Thus, if all in the above is correct, then the general on-shell component super particle actions, invariant under $S$ supersymmetry, must be of the form
\be\label{4}
S= \alpha \int dt + \int dt {\cal E} {\cal F}\left[ {\cal D}_t q {\cal D}_t  q \right],
\ee
where ${\cal F}$ is some arbitrary, for the time being, function, and $\alpha$ is a constant parameter. 
Funny enough, this function can be easily fixed by the bosonic action for the  particle
\be\label{5}
S_{bos}= \int dt \left( 1-\sqrt{1- \beta\dot{q}\dot{q}}\right),
\ee
where the value of the constant parameter $\beta$ is defined by the exact form of the superalgebra \p{1}.
Keeping in mind that in the bosonic limit
\be\label{6}
{\cal E}_{bos}=1, \qquad \left({\cal D}_t q\right)_{bos} = \dot{q},
\ee we conclude that the most general component action possessing the proper bosonic limit \p{5}  and invariant under
spontaneously broken supersymmetry has the form
\be\label{7}
S= \alpha \int dt + (1-\alpha) \int dt\, {\cal E} -\int dt\, {\cal E} \sqrt{1- \beta {\cal D}_t q {\cal D}_t{q}}\;.
\ee
The role of the unbroken $Q$ supersymmetry is just to fix the constant parameters $\alpha, \beta$ in the action \p{7}.
All differences between models with particular patterns of different supersymmetries breaking will be
in the concrete structure of the einbein ${\cal E}$ and covariant derivatives ${\cal D}_t q$, only.

The main goal of this paper is to check the validity of this statement for the super particles in $D=3$ and $D=5$.
In the next section we will present a detailed construction of the super particle action in $D=3$ realizing the 
$N=16 \rightarrow N=8$ pattern of supersymmetry breaking, with the chiral Goldstone supermultiplet. Then we will generalize the 
action to the $N=4\cdot 2^{k} \rightarrow N=2\cdot 2^k$ cases. In section 3 we will consider
the super particle action, again realizing the $N=16 \rightarrow N=8$ pattern of supersymmetry breaking, but in $D=5$.
All these explicit actions confirm our conjecture about the structure of the component action. We conclude with some comments and perspectives.

\section{Chiral supermultiplet}
The main goal of this section is to provide the detailed structure of the component on-shell actions for the
one dimensional systems realizing the 1/2 breaking of the global supersymmetry. We start with the edifying example of the system 
with the $N=16 \rightarrow N=8$ pattern of supersymmetry breaking and then will generalize
the constructed action to  the general $N=4\cdot 2^{k} \rightarrow N=2\cdot 2^k$  case.
\subsection{$N=16 \rightarrow N=8$ with chiral supermultiplet}
\subsubsection{Superfields Coset approach: kinematic}
It is a well known fact that the action for the given pattern of the supersymmetry breaking is completely defined by
the choice of the corresponding Goldstone supermultiplet \cite{BG1,BG2,BG3,RT,R2,IK,BIK1,BIK2,BIK3}. The bosonic scalars in 
the supermultiplet are associated with the central charges in the supersymmetry algebra.
Thus, for the system with the chiral supermultiplet one has to choose $N=16, d=1$ Poincar\'{e} superalgebra with two central charges:
\be\label{N16algebra}
\left\{ Q^{i a}, \overline Q_{j b}  \right\}=2\delta^a_b \delta^i_j P\,, \quad
\left\{ S^{i a}, \overline S_{j b}  \right\}=2\delta^a_b \delta^i_j P\,, \qquad
\left\{ Q^{ia}, S^{j b}  \right\} =2 i \eps^{a b} \eps^{ij} Z\,, \quad
\left\{ \overline Q_{i a}, \overline S_{j b}  \right\} = - 2 i \eps_{ab} \eps_{ij} \overline Z\,.
\ee
Here, $Q^{ia}, \bQ_{ia}$ and $S^{ia}, \bS_{ia}$ are the generators of unbroken and spontaneously broken $N=8$ supersymmetries, respectively. 
$P$ is the generator of translations, while $Z$ and $\bZ$ are  the central charge generators. The indices $i,a =1,2$ refer to the indices 
of the fundamental representations of the two commuting $SU(2)$ groups.

In the coset approach \cite{coset1,coset2}  the statement that the $S$ supersymmetry and $Z,\bZ$ translations are spontaneously broken 
is reflected in the structure of the group element $g$:
\be\label{cosetN16}
g=e^{i tP}\,e^{\theta_{i a} Q^{i a} + \bar \theta^{i a} \overline Q_{i a}}\, e^{i (\mq Z+\bar\mq \overline Z)}\,
e^{\mpsi_{i a} S^{i a} + \bar{\mpsi}^{i a} \overline S_{i a}}.
\ee
Once we state that the coordinates $\mpsi$ and $\mq$ are the superfields depending on the superspace
coordinates $\{ t, \theta,\bar\theta\}$, then we are dealing with the spontaneously breaking of the corresponding
symmetries. Thus, in our case we will treat $\mpsi(t,\theta,\bar\theta),\, \mq(t,\theta,\bar\theta)$ as $N=8$ Goldstone superfields 
accompanying the $N=16 \rightarrow N=8$ breaking of supersymmetry.

The transformation properties of the coordinates and superfields under the unbroken and broken supersymmetries
are induced by the left multiplications of the group element \p{cosetN16}:
$$g_0\, g = g'\, .$$
Thus, for the unbroken supersymmetry with $g_0= e^{\eps_{i a} Q^{i a} + \bar \eps^{i a} \overline Q_{i a}}$
one gets
\be\label{N16tran-U}
\delta_Q t = i \left(\eps_{i a} \bar \theta^{i a} + \bar \eps^{i a} \theta_{i a}  \right), \quad
\delta_Q\theta_{ia} = \eps_{i a}, \quad \delta_Q \bar \theta^{i a} = \bar \eps^{i a},
\ee
while for the broken supersymmetry, with $g_0=e^{\eta_{i a} S^{i a} + \bar{\eta}^{i a} \overline S_{i a}}$,
the transformations read
\be\label{N16tran-B}
\delta_S t = i \left(\eta_{i a} \bar \mpsi^{i a} + \bar \eta^{i a} \mpsi_{i a}  \right), \quad
\delta_S\mpsi_{i a} = \eta_{i a}, \quad \delta_S\bar \mpsi^{i a} = \bar \eta^{i a}, \quad
\delta_S \mq = - 2 \eta_{i a} \theta^{i a}, \quad \delta_S \bar \mq = 2 \bar \eta^{i a} \bar\theta_{i a}.
\ee

The local geometric properties of the system are specified by the left-invariant Cartan forms
\be\label{N16-Cartan}
g^{-1} dg = i \omega_P P + (\omega_{Q})_{i a}Q^{i a} + (\bar\omega_{Q})^{i a} \overline Q_{i a}
+ i \omega_Z Z + i \bar\omega_Z \overline Z
+ (\omega_{S})_{i a} S^{i a} +( \bar\omega_{S})^{i a} \overline S_{i a}
\ee
which look extremely simple in  our case:
\bea\label{N16-form}
&&\omega_P = dt -i (\bar \theta^{ia} d\theta_{ia}  + \theta_{ia} d \bar \theta^{ia}
+  \bar \mpsi^{ia} d\mpsi_{ia} + \mpsi_{ia} d \bar \mpsi^{ia}),\quad (\omega_Q)_{ia} = d\theta_{ia}, \quad (\bar\omega_{Q})^{ia} = d\bar \theta^{ia},\nn \\
&& (\omega_{S})_{ia} = d\mpsi_{ia},\quad (\bar\omega_{S})^{ia} = d\bar{\mpsi}^{ia},\quad
\omega_Z = d \mq + 2\mpsi^{ia} d\theta_{ia},\quad \bar \omega_Z = d \bar \mq - 2 \bar \mpsi_{ia} d \bar \theta^{ia}.
\eea
It is worth to note, that the all Cartan forms \p{N16-form} are invariant under the transformations \p{N16tran-U} and
\p{N16tran-B}.

Using the covariant differentials $\{ \omega_P, d\theta_{ia}, d \bar \theta^{ia}\}$ \p{N16-form},
one may construct the covariant derivatives
\bea\label{d1N16covder1}
&&
\partial_t = E \nabla_t \,, \quad E = 1- i \left( \mpsi_{ia} \dot{\bar\mpsi}{}^{ia}
+ \bar \mpsi^{ia} \dot \mpsi_{ia}  \right), \quad
E^{-1} = 1+ i \left( \mpsi_{ia} \nabla_t \bar \mpsi^{ia} + \bar \mpsi^{ia} \nabla_t \mpsi_{ia}  \right), \nn\\
&&
\nabla^{ia} = D^{ia} - i \left( \mpsi_{kb} D^{ia} \bar \mpsi^{kb} + \bar \mpsi^{kb} D^{ia} \mpsi_{kb} \right)\nabla_t
= D^{ia} - i \left( \mpsi_{kb} \nabla^{ia} \bar \mpsi^{kb} + \bar \mpsi^{kb} \nabla^{ia} \mpsi_{kb} \right)\partial_t, \nn \\
&&
\overline{\nabla}_{ia} = \overline{D}_{ia} - i \left( \mpsi_{kb} \overline{D}_{ia} \bar \mpsi^{kb}
+ \bar \mpsi^{kb} \overline{D}_{ia} \mpsi_{kb} \right)\nabla_t =
\overline{D}_{ia} - i \left( \mpsi_{kb} \overline{\nabla}_{ia} \bar \mpsi^{kb}
+ \bar \mpsi^{kb} \overline{\nabla}_{ia} \mpsi_{kb} \right)\partial_t\,,
\eea
where
\be\label{N16-flatDer}
D^{ia} = \frac{\partial}{\partial \theta_{i a}} -i \bar \theta^{i a} \partial_t, \quad
\overline D_{ia} = \frac{\partial}{\partial \bar \theta^{i a}} -i \theta_{i a} \partial_t,  \quad
\left\{ D^{ia} , \overline D_{jb} \right\}
= -2 i \delta^a_b \delta^i_j \partial_t.
\ee
These derivatives satisfy the following (anti)commutation relations
\bea\label{d1N16covder2}
&&
\left\{ \nabla^{ia} , \nabla^{jb} \right\}
= -2 i \left( \nabla^{ia} \mpsi_{kc} \nabla^{jb} \bar \mpsi^{kc}
+ \nabla^{ia} \bar \mpsi^{kc} \nabla^{jb} \mpsi_{kc} \right)\nabla_t,\nn \\
&&
\left\{ \overline{\nabla}_{ia} , \overline{\nabla}_{jb} \right\}
= -2 i \left( \overline{\nabla}_{ia} \mpsi_{kc} \overline{\nabla}_{jb} \bar \mpsi^{kc}
+ \overline{\nabla}_{ia} \bar \mpsi^{kc} \overline{\nabla}_{jb} \mpsi_{kc} \right)\nabla_t,\nn \\
&&
\left[ \nabla_t , \nabla^{ia} \right]
= -2 i \left( \nabla_t \mpsi_{kc} \nabla^{ia} \bar \mpsi^{kc}
+ \nabla_t \bar \mpsi^{kc}  \nabla^{ia} \mpsi_{kc} \right)\nabla_t, \nn \\
&&
\left[ \nabla_t , \overline{\nabla}_{ia} \right]
= -2 i \left( \nabla_t \mpsi_{kc} \overline{\nabla}_{ia} \bar \mpsi^{kc}
+ \nabla_t \bar \mpsi^{kc}  \overline{\nabla}_{ia} \mpsi_{kc} \right)\nabla_t, \nn \\
&&
\left\{ \nabla^{ia} , \overline{\nabla}_{jb} \right\}
= -2 i \delta^a_b \delta^i_j \nabla_t -2 i \left( \nabla^{ia} \mpsi_{kc} \overline{\nabla}_{jb} \bar \mpsi^{kc}
+ \nabla^{ia} \bar \mpsi^{kc} \overline{\nabla}_{jb} \mpsi_{kc}  \right)\nabla_t.
\eea

Finally, one may reduce the number of independent Goldstone superfields by imposing the conditions on the 
$d\theta$-projections of the Cartan forms $\omega_Z, \bar \omega_Z$ \p{N16-form}
\be\label{N16-kin}
\left\{
\begin{array}{l}
\omega{}_Z|_{\theta} =0, \\
\overline{\omega}_Z|_{\theta} = 0,
\end{array}\right.
\quad \Rightarrow \quad \left\{
\begin{array}{l}
\overline{\nabla}_{i a} \mq =0, \quad \nabla^{i a} \mq - 2 \mpsi^{i a} =0,\\
\nabla^{i a} \bar \mq =0, \quad \overline{\nabla}_{i a} \bar \mq + 2 \bar \mpsi_{i a} =0.
\end{array}\right.
\ee
These constraints are purely kinematical ones. They impose the covariant chirality conditions on the superfields
$\mq$ and $\bar\mq$, and in addition they express the fermionic Goldstone superfields $\mpsi^{i a}, \bar \mpsi_{i a}$
as the spinor derivatives of the $\mq$ and $\bar\mq$, thereby realizing the Inverse Higgs phenomenon \cite{ih}.

Thus, in order to realize the $N=16 \rightarrow N=8$ breaking of the global supersymmetry in one dimension we need one,
covariantly chiral, $N=8$ bosonic superfield $\mq(t,\theta,\bar\theta)$.

\subsubsection{Superfields Coset approach: dynamics}
It is well known that the chirality conditions are not enough to select an irreducible $N=8$ supermultiplet: one has to impose
additional, second order in the spinor derivatives constraints on the superfield $\{ \mq,\bar\mq \}$ \cite{BIKL1}.
Unfortunately, as it often happened in the coset approach,  the direct covariantization of the irreducibility constraints is 
not covariant \cite{BG2}, while the simultaneous covariantization of the constraints and the equations of motion works perfectly. 
That is why we propose the following equations which should describe our super particle:
\be\label{N16-dynamics}
\nabla^{i a} \mpsi_{j b} = 0, \qquad \overline{\nabla}_{i a} \bar \mpsi^{j b} =0.
\ee
These equations are covariant with respect to both unbroken and broken supersymmetries. Moreover, one may easily find that in 
the bosonic limit they amount to the following equation of motion for the scalar field $q=\mq|_{\theta=0}$:
\be\label{beom}
\frac{d}{dt}\left[\frac{\dot{q}}{\sqrt{1- 4 \dot{q} \dot{\bar q}}}\right]=0.
\ee
The equation \p{beom} follows from the bosonic action
\be\label{bS}
S_{bos}= \int dt \left( 1 - \sqrt{1-4 \dot{q} \dot{\bar q}}\right)
\ee
which is a proper action for a particle in $D=3$ space-time.

At this point one should wonder whether the equations \p{N16-dynamics} are self-consistent. Indeed, due to
eqs. \p{N16-kin} from \p{N16-dynamics} we have
\be\label{selfcon}
\nabla^{i a} \mpsi_{j b} =\frac{1}{2} \nabla^{ia} \nabla_{jb} \mq =0 \qquad \Rightarrow \qquad \{ \nabla^{ia},\nabla_{jb}\} \mq=0 \;.
\ee
So, one may expect some additional conditions on the superfield $\mq$ due to the relations \p{d1N16covder2}.
However, on the constraints surface in \p{N16-dynamics} we have
\be\label{selfcon1}
\left\{ \nabla^{ia} , \nabla^{jb} \right\}=0, \qquad
\left\{ \overline{\nabla}_{ia} , \overline{\nabla}_{jb} \right\}= 0
\ee
and thus the equations \p{N16-dynamics} are perfectly self-consistent.

One should note that the rest of the commutators in \p{d1N16covder2} are also simplified as
\be\label{CDalgebra}
\left \{\nabla^{i a} , \overline{\nabla}_{j b} \right \} = - 2i \delta^i_j \delta^a_b (1+ \mlambda \bar \mlambda)\nabla_t, \quad
\left [\nabla_t, \nabla^{i a} \right ] = 2i \bar \mlambda \nabla_t \mpsi^{i a} \nabla_t,\quad
\left [ \nabla_t, \overline{\nabla}_{i a} \right ] = 2i \mlambda \nabla_t \bar{\mpsi}_{i a} \nabla_t,
\ee
where we introduced the superfields $\{ \mlambda, \bar\mlambda \}$
\be\label{lambda}
\left\{
\begin{array}{l}
\overline{\nabla}_{i a} \mpsi_{j b} + \eps_{ij} \eps_{a b} \mlambda =0,\\
\nabla^{i a} \bar \mpsi^{j b} + \eps^{ij} \eps^{a b} \bar \mlambda =0.
\end{array}\right.
\quad \Rightarrow \quad
\left\{
\begin{array}{l}
\nabla_t \mq +\frac{i \mlambda}{1+ \mlambda \bar \mlambda} = 0\,, \\
\nabla_t \bar \mq - \frac{i \bar \mlambda}{1+ \mlambda \bar \mlambda} = 0.
\end{array}\right.
\ee
\subsubsection{Components Coset approach}
Despite the explicit construction of the proper equations of motion within the superfields version of the coset approach, the latter
is poorly adapted for the construction of the action. That is why in the paper \cite{BKS1} the component version of the coset  
approach  has been proposed, in order to construct the actions. In the application to the present case, the basic steps of this 
method can be formulated as follows:
\begin{itemize}
\item Firstly, on-shell our $N=8$ supermultiplet $\{ \mq, \bar\mq \}$ contains the following physical components:
$$ q=\mq|_{\theta=0}, \quad \bar q= \bar \mq|_{\theta=0}, \quad \psi_{ia}=\mpsi_{ia}|_{\theta=0}, \quad \bpsi^{ia}= \bar\mpsi^{ia}|_{\theta=0}.$$
They are just the first components of the superfield parameterizing the coset \p{cosetN16}.
\item Secondly, with respect to the broken supersymmetry $\delta\theta=\delta\bar\theta=0$ \p{N16tran-B}. This means, that the 
transformation properties of the physical components $\{ q, \bar q, \psi,\bar\psi \}$ under the broken supersymmetry can be extracted from the coset
    \be\label{cosetVA}
g|_{\theta=0}=e^{i tP}\, e^{i (q Z+\bar{q} \overline Z)}\,
e^{\psi_{i a} S^{i a} + \bar{\psi}^{i a} \overline S_{i a}}.
\ee
In other words, the fields $\{q,{\bar q}, \psi_{ia}, \bpsi^{ia}\}$ parameterize the coset \p{cosetVA} which is responsible for 
the full breaking of the $S$ supersymmetry. Moreover, with respect to this supersymmetry the fields $ q,{\bar q}$ are 
just matter fields, because $\delta_S q= \delta_S {\bar q}=0$, while the fermions $\psi_{ia},\bpsi^{ia}$ are just Goldstone fermions. 
This means that the component action has to be of the Volkov-Akulov type \cite{VA}, i.e. the fermions $\psi_{ia},\bpsi^{ia}$ may 
enter the action through the einbein $\cal E$ or through
the covariant derivatives ${\cal D}_t q, {\cal D}_t{\bar q}$ only, with
\be\label{cov1}
\partial_t={\cal E}{\cal D}_t, \quad {\cal E} = E|_{\theta=0}=1- i \left( \psi_{ia} \dot{\bar\psi}{}^{ia}
+ \bar \psi^{ia} \dot \psi_{ia}  \right), \quad
{\cal E}^{-1} = 1+ i \left( \psi_{ia} {\cal D}_t \bar \psi^{ia} + \bar \psi^{ia} {\cal D}_t \psi_{ia}  \right).
\ee

Thus, the unique candidate to be the component on-shell action, invariant with respect to spontaneously broken $S$ supersymmetry  reads
\be\label{action1}
S= \alpha \int dt + \int dt {\cal E} {\cal F}\left[ {\cal D}_t q {\cal D}_t{\bar q}\right]
\ee
with the arbitrary, for the time being, function $\cal F$ and constant parameter $\alpha$.
\item Finally, considering the bosonic limit of the action \p{action1} and comparing it with the known
bosonic action \p{bS} one may find the function ${\cal F}$:
\be\label{F}
\int dt \left( \alpha +  {\cal F}\left[\dot q \bar{\dot q}\right]\right) 
=\int dt\left(1-\sqrt{1-4 \dot{q} \dot{\bar q}}\right)\; \Rightarrow\; {\cal F} =\left(1-\alpha -\sqrt{1-4 \dot{q} \dot{\bar q}}\right)\;.
\ee
\end{itemize}
Therefore, the most general component action possessing the proper bosonic limit \p{bS} and invariant under
spontaneously broken supersymmetry has the form
\be\label{action2}
S= \alpha \int dt + (1-\alpha) \int dt\, {\cal E} -\int dt\, {\cal E} \sqrt{1- 4 {\cal D}_t q {\cal D}_t{\bar q}}\;.
\ee

In principle, the invariance of the action \p{action2} under broken supersymmetry is evident. Nevertheless,
for completeness, let us demonstrate it explicitly.

From \p{N16tran-B} we know the total variations of our components and the time $t$:
\be\label{invB1}
\delta_S t = i \left(\eta_{i a} \bar \psi^{i a} + \bar \eta^{i a} \psi_{i a}  \right), \quad
\delta_S\psi_{i a} = \eta_{i a}, \quad \delta_S\bar \psi^{i a} = \bar \eta^{i a}, \quad
\delta_S q =0, \quad \delta_S \bar q = 0.
\ee
Therefore, the transformations of the components in the fixed point read
\be\label{inB2}
\delta^*_S q = \delta_S q - \delta_S t \dot q , \quad \delta^*_S \psi_{ia} = \delta_S \psi_{ia} - \delta_S t \dot{\psi_{ia}}.
\ee
It immediately follows from \p{inB2} and definitions \p{cov1} that
\be\label{invB3}
\delta^*_S\left( {\cal E} {\cal F}\left[ {\cal D}_t q {\cal D}_t{\bar q}\right]\right)=
-i \partial_t \left[ \left( \eta_{i a} \bar \psi^{i a} + \bar \eta^{i a} \psi_{i a}  \right) {\cal E} {\cal F}\left[ {\cal D}_t q {\cal D}_t{\bar q}\right]\right].
\ee
Thus, two last terms in the action \p{action2} are invariant, while the invariance of the first, trivial term is evident.

The final step is to check the invariance of the action \p{action2} under the unbroken $Q$ supersymmetry which is
realized on the components as follows
\bea\label{Ususy1}
&& \delta^*_Q q = -2 \eps^{i a}  \psi_{i a} + i \left( \eps^{i a} \psi_{i a}\bar\lambda+
\bar\eps^{i a}\bpsi_{i a} \lambda\right) \partial_t q \nn \\
&& \delta^*_Q \psi_{i a} = \bar\eps_{i a}  \lambda + i \left( \eps^{j b} \psi_{j b}\bar\lambda+
\bar\eps^{j b}\bpsi_{j b} \lambda\right) \partial_t \psi_{i a} \;.
\eea
Here, $\lambda$ is the first component of the superfield $\mlambda$ defined in \p{lambda}
\be
\lambda = \frac{2 i {\cal D}_t q}{1+\sqrt{1- 4 {\cal D}_t q {\cal D}_t{\bar q}}}\,.
\ee
From \p{Ususy1} and the definitions \p{cov1} one may easily find the transformation properties of the main ingredients
\bea\label{Ususy2}
&& \delta^*_Q {\cal E} = i \partial_t \left[ \left( \eps^{j b}\psi_{j b}  \bar\lambda 
+ \bar\eps^{j b}\bpsi_{j b}  \lambda \right) {\cal E}   \right] -2 i \big ( \eps^{j b}\dot\psi_{j b}  \bar\lambda 
+ \bar\eps^{j b}\dot\bpsi_{j b}  \lambda  \big ), \nn \\
&& \delta^*_Q {\cal D}_t q = i \left(  \eps^{j b}\psi_{j b}  \bar\lambda + \bar\eps^{j b}\bpsi_{j b}  \lambda   \right) \partial_t ({\cal D}_t q) 
-2\eps^{j b}{\cal D}_t \psi_{j b} + 2 i \left( \eps^{j b}{\cal D}_t\psi_{j b}  \bar\lambda + \bar\eps^{j b}{\cal D}_t\bpsi_{j b}  \lambda  \right){\cal D}_t q.
\eea
Now, one may calculate the variation of the integrand in the action \p{action1}
\bea\label{Ususy3}
&&\delta^*_Q \left({\cal  E}\; {\cal F}  \right) = 2\partial_t \left[ {\cal E} \frac{\eps^{j b}\psi_{j b} {\cal D}_t \bar q 
- \bar\eps^{j b}\bpsi_{j b}  {\cal D}_t q }{1+\sqrt{1-4{\cal D}_t q  {\cal D}_t \bar q}} {\cal F} \right]+ \nn \\
&&+ \frac{\eps^{j b}\dot\psi_{j b} {\cal D}_t \bar q - \bar\eps^{j b}\dot\bpsi_{j b}  
{\cal D}_t q}{1+\sqrt{1-4{\cal D}_t q {\cal D}_t \bar q}} \left[ -4 {\cal F} -2{\cal F}^\prime 
\left( 1+\sqrt{1-4{\cal D}_t q {\cal D}_t \bar q} - 4{\cal D}_t q {\cal D}_t \bar q \right)  \right].
\eea
Substituting the function ${\cal F}$ \p{F} and its derivative over its argument ${\cal D}_t q {\cal D}_t \bar q$, we 
will find that the second term in the variation \p{Ususy3} is canceled if $\alpha=2$.
Keeping in the mind that the first term in the action \p{action2} is trivially invariant under unbroken supersymmetry,
we conclude that the unique component action, invariant under both unbroken $Q$ and broken $S$ $N=8$ supersymmetries reads
\be\label{action3}
S= 2 \int dt -  \int dt\, {\cal E} \left(1 + \sqrt{1- 4 {\cal D}_t q {\cal D}_t{\bar q}}\right)\;.
\ee
\subsection{Rolling down}
The construction of the component action, we considered in the previous section, has two interesting peculiarities:
\begin{itemize}
\item It is based on the coset realization of the $N=16$ superalgebra \p{N16algebra};
\item In the component action \p{action3} the summation over indices $\{i,a\}$ of two $SU(2)$ groups affected
only physical fermions $\{ \psi_{ia}, \bpsi^{ia} \}$.
\end{itemize}
It is quite clear, that in such a situation one may consider two subalgebras of the $N=16$ superalgebra
\begin{itemize}
\item $N=8$ supersymmetry, by choosing the corresponding supercharges as
\be\label{N8} {\tilde Q}^{i} \equiv Q^{i1},\; {\widetilde\bQ}_{i} \equiv \bQ_{i1}, \quad {\tilde S}^{i} \equiv S^{i2},\; {\widetilde\bS}_{i} \equiv \bS_{i2}
\ee
\item $N=4$ supersymmetry with the supercharges
\be\label{N4} {\hat Q} \equiv Q^{11},\; {\widehat\bQ} \equiv \bQ_{11}, \quad {\hat S}\equiv S^{22},\; {\widehat\bS} \equiv \bS_{22}.
\ee
\end{itemize}
It is evident that the corresponding component actions will be given by the same expression \p{action3}, in which
the ``new'' einbeins and covariant derivatives read
\be\label{covN8}\mbox{ $N=8$ case: } \left\{
\partial_t=\tilde{\cal E}\tilde{\cal D}_t, \quad \tilde{\cal E} = 1- i \left( \psi_{i2} \dot{\bar\psi}{}^{i2}
+ \bar \psi^{i2} \dot \psi_{i2}  \right), \quad
\tilde{\cal E}^{-1} = 1+ i \left( \psi_{i2} \tilde{\cal D}_t \bar \psi^{i2} + \bar \psi^{i2} \tilde{\cal D}_t \psi_{i2}  \right),
\right.
\ee
\be\label{covN4}\mbox{ $N=4$ case: } \left\{
\partial_t=\hat{\cal E}\hat{\cal D}_t, \quad \hat{\cal E} = 1- i \left( \psi_{22} \dot{\bar\psi}{}^{22}
+ \bar \psi^{22} \dot \psi_{22}  \right), \quad
\hat{\cal E}^{-1} = 1+ i \left( \psi_{22} \hat{\cal D}_t \bar \psi^{22} + \bar \psi^{22} \hat{\cal D}_t \psi_{22}  \right).
\right.
\ee
Thus, we see that the action \p{action3} has a universal character, describing the series of theories with
the following patterns of global supersymmetry breaking:
$N=16\rightarrow N=8$, $N=8\rightarrow N=4$ and $N=4\rightarrow N=2$.
\subsection{Climbing up}
It is almost evident, that the universality of the action \p{action3} can be used to extend our construction
to the cases of $N=4\cdot 2^k$ supersymmetries by adding the needed numbers of $SU(2)$ indices to the superscharges as
\be\label{genSUSY1}
Q \rightarrow Q^{\alpha_1\ldots\alpha_k},\; \bQ \rightarrow \bQ_{\alpha_1\ldots\alpha_k},\quad S \rightarrow 
S^{\alpha_1\ldots\alpha_k},\; \bS \rightarrow \bS_{\alpha_1\ldots\alpha_k},
\ee
obeying the $N=4\cdot 2^k$ Poincar\'e superalgebra
\bea\label{genalgebra}
&& \left\{ Q^{\alpha_1\ldots\alpha_k}, \overline Q_{\beta_1\ldots\beta_k}  \right\}=2\delta^{\alpha_1}_{\beta_1}\ldots \delta^{\alpha_k}_{\beta_k} P\,, \quad
\left\{ S^{\alpha_1\ldots\alpha_k}, \overline S_{\beta_1\ldots\beta_k}  \right\}=2\delta^{\alpha_1}_{\beta_1}\ldots \delta^{\alpha_k}_{\beta_k} P\,,\nn \\
&&\left\{ Q^{\alpha_1\ldots\alpha_k}, S^{\beta_1\ldots\beta_k}  \right\}=2 i \eps^{\alpha_1 \beta_1}\ldots \eps^{\alpha_k \beta_k} Z\,, \quad
\left\{ \overline Q_{\alpha_1\ldots\alpha_k}, \overline S_{\beta_1\ldots\beta_k}  \right\}=-2 i \eps_{\alpha_1 \beta_1}\ldots \eps_{\alpha_k\beta_k} \overline Z \,.
\eea
Once again, the component action describing super particles in $D=3$ space with $N=4\cdot 2^k$ Poincar\'e supersymmetry
partially broken down to the $N=2\cdot 2^k$ one will be given by the same expression \p{action3} with the following substitutions:
\be\label{genpsi}
\psi \rightarrow \psi_{\alpha_1\ldots\alpha_k},\quad \bpsi \rightarrow \bpsi^{\alpha_1\ldots\alpha_k},\quad
{\cal E} = 1- i \left( \psi_{\alpha_1\ldots\alpha_k} \dot{\bar\psi}{}^{\alpha_1\ldots\alpha_k}
+ \bar \psi^{\alpha_1\ldots\alpha_k} \dot \psi_{\alpha_1\ldots\alpha_k}  \right).
\ee
\setcounter{equation}0
\section{Super particle in $D=5$}
In this section we will apply our approach to the $N=16$ super particle in $D=5$. The corresponding superfield 
equations of motion for this system, which possesses 8 manifest and 8 spontaneously broken supersymmetries, have been constructed
within the coset approach in \cite{BIK2}, while the action is still unknown.

In order to describe the super particle in $D=5$ with 16 supersymmetries one has to start with the following superalgebra
\be\label{algebra}
\{Q^i_{\alpha}, Q^j_{\beta} \} = \eps^{ij} \Omega_{\alpha \beta} P, \quad
\{Q^i_{\alpha}, S^{b \beta} \} = \delta^{\beta}_{\alpha} Z^{ib}, \quad
\{S^{a \alpha}, S^{b \beta} \} = - \eps^{ab} \Omega^{\alpha \beta} P,\quad (i,a=1,2; \alpha,\beta =1,2,3,4)
\ee
where the invariant $Spin(5)$ symplectic metric $\Omega_{\alpha\beta}$, allowing to raise and lower the spinor indices,
obeys the conditions\footnote{We use the following convention: $\eps^{\alpha \beta \lambda \sigma} \eps_{\alpha \beta \lambda \sigma} = 24\,,\;
\eps^{\alpha \beta \lambda \sigma}\eps_{\alpha \beta \mu \rho} = 2 (\delta^{\lambda}_{\mu}\, \delta^{\sigma}_{\rho} -
\delta^{\lambda}_{\rho}\, \delta^{\sigma}_{\mu})$}
\bea\label{Omega}
&&
\Omega_{\alpha \beta} = - \Omega_{\beta \alpha}\,, \quad
\Omega^{\alpha \beta} = - \frac{1}{2}\, \eps^{\alpha \beta \lambda \sigma} \Omega_{\lambda \sigma}\,, \quad
\Omega_{\alpha \beta} = - \frac{1}{2}\, \eps_{\alpha \beta \lambda \sigma} \Omega^{\lambda \sigma}\,, \quad
\Omega_{\alpha \beta} \Omega^{\beta \gamma} = \delta_{\alpha}^{\gamma}\,.
\eea
From the one-dimensional perspective this algebra is the $N=16$ super Poincar\'{e} algebra with four central charges $Z^{ia}$.
If we are going to treat $S$ supersymmetry to be spontaneously broken, than we have to consider the following element 
of the coset:\footnote{Here, we strictly follow the notations adopted in \cite{BIK2} which are slightly different from those we used in the previous Sections.}
\be\label{coset}
g = e^{tP}\, e^{\theta_i^{\alpha} Q^i_{\alpha}}\, e^{\mq_{ia} Z^{ia}}\,
e^{\mpsi_{a \alpha} S^{a \alpha}}.
\ee
Here $(t, \theta_i^{\alpha})$ are the coordinates of $N=8, d=1$ superspace while $\mq_{ia} = \mq_{ia}(t,\theta_i^{\alpha}),$
$\mpsi_{a \alpha}= \mpsi_{a \alpha}(t,\theta_i^{\alpha}),$  are Goldstone superfields.

Similarly to the cases we considered in the previous sections, one may find the transformation properties of the 
coordinates and superfields, by acting from the left on the coset element \p{coset}
by different elements of the group with constant parameters.
So, for the unbroken supersymmetry ($g_0=\exp{(\eps^{\alpha}_i Q_{\alpha}^i)}$) one gets
\be\label{U-susy}
\delta_Q t = - \frac{1}{2}\,\eps^{\alpha}_i \theta^{i \beta} \Omega_{\alpha \beta}\,, \quad
\delta_Q \theta^{\alpha}_i = \eps^{\alpha}_i\,,
\ee
while for the broken supersymmetry ($g_0=\exp{(\eta_{a \alpha} S^{a \alpha})}$) the corresponding transformations read
\be\label{B-susy}
\delta_S t = - \frac{1}{2}\, \eta^a_{\alpha} \mpsi_{a \beta} \Omega^{\alpha \beta}\,, \quad
\delta_S \mpsi_{a \alpha} = \eta_{a \alpha}\,, \quad
\delta_S \mq_{ia} = - \eta_{a \alpha}\theta^{\alpha}_i\,.
\ee
The last ingredient we need is the Cartan forms, defined in a standard way as
\be\label{full-Cartan}
g^{-1} dg = \omega_P P + (\omega_Q)_i^{\alpha}\, Q^i_{\alpha} + (\omega_Z)_{ia}\, Z^{ia}
+ (\omega_S)_{a \alpha}\, S^{a \alpha}\,,
\ee
with
\bea\label{Cartan}
&& \omega_P= dt - \frac{1}{2}\left( d \theta_i^{\alpha} \theta^{i \beta} +
 d \mpsi_a^{\alpha} \mpsi^{a \beta}\right) \Omega_{\alpha \beta},\quad
 (\omega_Z)_{ia} = d\mq_{ia} -  d \theta^{\alpha}_i \mpsi_{a \alpha}, \nn \\
&& (\omega_Q)_i^{\alpha} =  d \theta^{\alpha}_i, \quad
(\omega_S)_{a \alpha} = d \mpsi_{a \alpha}.
\eea
Using the covariant differentials $\{\omega_P,(\omega_Q)_i^{\alpha}\}$ one may construct the covariant derivatives $\nabla_t$ and $\nabla^i_{\alpha}$
\bea\label{cov-der}
\partial_t &=& E\, \nabla_t\,, \quad
E = 1 + \frac{1}{2}\, \Omega^{\beta \gamma} \mpsi^a_{\beta} \partial_t \mpsi_{a \gamma}\,, \quad
E^{-1} = 1 - \frac{1}{2}\, \Omega^{\beta \gamma} \mpsi^a_{\beta} \nabla_t \mpsi_{a \gamma}\,,\\
\nabla^i_{\alpha} &=& D^i_{\alpha} + \frac{1}{2}\, \Omega^{\beta \gamma} \mpsi^a_{\beta}
D^i_{\alpha} \mpsi_{a \gamma} \nabla_t
= D^i_{\alpha} + \frac{1}{2}\, \Omega^{\beta \gamma} \mpsi^a_{\beta}
\nabla^i_{\alpha} \mpsi_{a \gamma} \partial_t\,,
\eea
where
\be\label{flat-der1}
D^i_{\alpha} = \frac{\partial}{\partial \theta^{\alpha}_i}
+\frac{1}{2}\, \theta^{i \beta} \Omega_{\alpha \beta} \partial_t\,, \quad
\Big \{D^i_{\alpha}, D^j_{\beta} \Big \} = \eps^{ij}\, \Omega_{\alpha \beta}\, \partial_t\,.
\ee
These covariant derivatives satisfy the following (anti)commutation relations
\bea\label{relations-der}
&&
\Big \{\nabla^i_{\alpha}, \nabla^j_{\beta} \Big \} = \eps^{ij}\, \Omega_{\alpha \beta}\,
\nabla_t + \Omega^{\lambda \sigma}\, \nabla^i_{\alpha} \mpsi^b_{\lambda}\,
\nabla^j_{\beta} \mpsi_{b \sigma}\, \nabla_t\,,\nn \\
&&
\Big [\nabla_t, \nabla^i_{\alpha} \Big ] = \Omega^{\beta \gamma}\, \nabla_t \mpsi^b_{\beta}\,
\nabla^i_{\alpha} \mpsi_{b \gamma} \nabla_t\,.
\eea

Now, in a full analogy with the previously considered cases, we impose the following invariant condition on the $d\theta$-projections 
of the Cartan form $(\omega_Z)_{ia}$  \p{Cartan}:
\be
(\omega_Z)_{ia}|_\theta=0 \quad \Rightarrow \quad \left\{
\begin{array}{l}
\nabla^{(j}_{\alpha}\, \mq^{i)}_a = 0\,,  \qquad \qquad \quad \quad \quad \;\quad \quad\; \mbox{(a)}\\
\nabla^i_{\alpha}\, \mq_{ia} - 2 \mpsi_{a \alpha}=0.\;\;\; \qquad \qquad \quad \quad \mbox{(b)}
\end{array}\right. \label{EoM-Zt}
\ee
The condition $(\ref{EoM-Zt}b)$ identifies the fermionic superfield $\mpsi_{a\alpha}$ with the spinor derivatives of the 
superfield $\mq_{ia}$, just reducing the independent superfields to bosonic $\mq_{ia}$ ones (this is once again the 
Inverse Higgs phenomenon \cite{ih}). The conditions $(\ref{EoM-Zt}a)$ are more restrictive - they nullify all auxiliary components
in the superfield $\mq_{ia}$. Indeed, it immediately follows from \p{EoM-Zt} that
\be\label{dopeq1}
\frac{3}{2} \nabla_\beta^j \mpsi_{a\alpha} = \left\{ \nabla_\beta^j, \nabla_\alpha^i\right\}\mq_{ia} -
\frac{1}{2} \left\{ \nabla_\alpha^j, \nabla_\beta^i\right\}\mq_{ia}.
\ee
Using anti-commutators \p{relations-der}, one may solve this equation as follows:
\be\label{EoM-Stt}
\nabla^j_{\beta}\, \mpsi_{a \alpha} + \frac{1}{2}\, \mlambda^j_a \Omega_{\alpha \beta} = 0\,,
\ee
where the superfield $\mlambda^j_a$ is defined as
\be\label{lambdaB}
\nabla_t\, \mq^{ia} - \frac{1}{2}\, \frac{\mlambda^{ia}}{1+\frac{\mlambda^2}{8}} = 0.
\ee
Thus, we have an on-shell situation. In \cite{BIK2} the corresponding bosonic equation of motion has been found to be
\be\label{boseom}
\frac{d}{dt} \left( \frac{\dot q_{ia}}{\sqrt{1-2 \dot q^{jb} \dot q_{jb}}}\right)=0,
\ee
where $q_{ia}=\mq_{ia}|_{\theta=0}$ are the first components of the superfield $\mq_{ia}$. The equation of motion 
\p{boseom} corresponds to the static-gauge form of the Nambu-Goto action for the massive particle in $D=5$ space-time
\be\label{NG5}
S_{bos} \sim \int dt\left(1- \sqrt{1- 2 \dot q^{ia} \dot q_{ia}}\right) .
\ee

To construct the on-shell, component action we will follow the same procedure which we described in
full details in subsection (2.1.3). So, we will omit unessential details concentrating only on the new
features.

If we are interested in the invariance with respect to the broken $S$ supersymmetry, then we may consider the reduced
coset element
\be\label{coset_red}
g|_{\theta=0} = e^{tP}\,  e^{q_{ia} Z^{ia}}\,
e^{\psi_{a \alpha} S^{a \alpha}}.
\ee
Here, $q_{ia}$ and $\psi_{a\alpha}$ are the first components of the superfields $\mq_{ia}$ and
$\mpsi_{a\alpha}$. Similarly to the discussion in subsection (2.1.3), the Goldstone fermions $\psi_{a\alpha}$
may enter the component action only through the einbein $\cal E$ and the covariant derivatives ${\cal D}_t q_{ia}$, defined as
\be\label{cd4c}
\partial_t = {\cal E}\, {\cal D}_t\,, \quad
{\cal E} = 1 + \frac{1}{2}\, \Omega^{\beta \gamma} \psi^a_{\beta} \partial_t \psi_{a \gamma}\,, \quad
{\cal E}^{-1} = 1 - \frac{1}{2}\, \Omega^{\beta \gamma} \psi^a_{\beta} {\cal D}_t \psi_{a \gamma}\,,
\ee
Keeping in the mind the known bosonic limit of the action \p{NG5}, we come to the unique candidate for the component on-shell action
\be\label{action24}
S= \alpha \int dt + (1-\alpha) \int dt\, {\cal E} -\int dt\, {\cal E} \sqrt{1- 2 {\cal D}_t q^{ia} {\cal D}_t q_{ia}}\;.
\ee
This action is perfectly invariant with respect to the broken $S$ supersymmetry, realized on the physical components and
their derivatives as
\be\label{susy-B1}
\delta_S^* q_{ia} = \frac{1}{2}\, \eta^b_{\alpha} \psi_{b \beta} \Omega^{\alpha \beta}
\partial_t q_{ia}\,, \quad
\delta_S^* (\cD_t\,q_{ia}) = \frac{1}{2}\, \eta^b_{\alpha} \psi_{b \beta} \Omega^{\alpha \beta} \partial_t (\cD_t\,q_{ia})\,,
\delta_S^* \psi_{a \alpha} = \eta_{a \alpha} + \frac{1}{2}\, \eta^b_{\beta} \psi_{b \lambda} \Omega^{\beta \lambda}
\partial_t \psi_{a \alpha}\,.
\ee
From \p{susy-B1} one may find the transformation properties of the einbein ${\cal E}$
\be\label{var-E-B}
\delta_S^* {\cal E} = \frac{1}{2}\,\eta^a_{\alpha} \partial_t \left( {\cal E} \Omega^{\alpha \beta} \psi_{a \beta}\right )\,.
\ee
Now, combining \p{susy-B1}  and \p{var-E-B}, we will get
\be\label{var1gen}
\delta_S^*\left( {\cal E} {\cal F}\left[ \cD_t\,q^{jb}\cD_t\,q_{jb}\right]\right) =
\frac{1}{2}\,\eta^a_{\alpha} \partial_t \left( \Omega^{\alpha \beta} \psi_{a \beta} \; {\cal E}\; {\cal F}\left[ \cD_t\,q^{jb}\cD_t\,q_{jb}\right]\right )\,,
\ee
and, therefore, the second and the third terms in the action \p{action24} are separately invariant with respect to 
$S$ supersymmetry. The first term in \p{action24} is trivially invariant with respect to both, broken and unbroken supersymmetries.

The last step is to impose the invariance with respect to the unbroken $Q$ supersymmetry. Under the transformations 
of unbroken supersymmetry taken in the fixed point the variation of any superfield reads
$$\delta^*_Q {\mathbf{F}}= - \eps^{\alpha}_i Q^i_{\alpha}{\mathbf{F}}\,.$$
From this one may find the variations of the components $q_{ia}$ and $\psi_{a \alpha}$
and their covariant derivatives:
\bea\label{susy-U1}
&&
\delta_Q^* q_{ia} = - \eps^{\alpha}_i \psi_{a \alpha}
+ \frac{1}{4}\, \eps^{\alpha}_j \lambda^{jb} \psi_{b \alpha} \partial_t q_{ia}\,,\nn \\
&&
\delta_Q^* (\cD_t q_{ia}) = - \eps^{\alpha}_i \cD_t \psi_{a \alpha}
+ \frac{1}{4}\, \eps^{\alpha}_j  \frac{\lambda_{ia}}{1+\frac{1}{8}\,\lambda^2}\,
\lambda^{jb}\cD_t \psi_{b \alpha}
+ \frac{1}{4}\,\eps^{\alpha}_j \lambda^{jb} \psi_{b \alpha} \partial_t (\cD_t q_{ia})\,, \nn \\
&&\label{susy-U2}
\delta_Q^* \psi_{a \alpha} = \frac{1}{2}\, \eps^{\beta}_j \Omega_{\alpha \beta} \lambda^j_a
+ \frac{1}{4}\, \eps^{\beta}_j \lambda^{jb} \psi_{b \beta} \partial_t \psi_{a \alpha}\,.
\eea
The variation of the ein-bein $E$ can be also computed and it reads
\be\label{var-E-U}
\delta_Q^* {\cal E} = \frac{1}{4}\,\eps^{\beta}_j \partial_t \left ( {\cal E} \lambda^{j b} \psi_{b \beta} \right )
-  \frac{1}{2}\,\eps^{\beta}_j \lambda^{j b} \partial_t \psi_{b \beta}\,.
\ee
It is a matter of lengthly, but straightforward calculations to check, that the action \p{action24} is invariant under
the unbroken supersymmetry \p{susy-U2}, \p{var-E-U} if $\alpha=2$.

Thus, the component action, invariant under both unbroken $Q$ and broken $S$ $N=8$ supersymmetries reads
\be\label{N16-actionZZ}
S = \int dt \left [2-{\cal E}\, \left ( 1+ \sqrt{1-2 \cD_t q^{ia} \cD_t q_{ia}}\right )
\right ]\,.
\ee

\section{Conclusion}
In this paper we have checked  the validity of the conjecture that the on-shell component  super particle actions have the universal form
$$
S= \alpha \int dt + (1-\alpha) \int dt\, {\cal E} -\int dt\, {\cal E} \sqrt{1- \beta {\cal D}_t q {\cal D}_t{q}}\;.
$$
We explicitly constructed such actions for the super particles in $D=3$ realizing the $N=4\cdot 2^{k} \rightarrow N=2\cdot 2^k$ 
pattern of supersymmetry breaking, and in $D=5$ with the $N=16$ supersymmetry broken down to the $N=8$ one.
All constructed actions have indeed the universal form, confirming our conjecture.

Of course, the particular examples, we considered in the present paper, cannot replace the rigorous proof, but the details of 
calculations, where almost nothing depends on the number of broken supersymmetries, almost convinced us that our conjecture is correct.

The most important features of our construction may be summarized as follows:
\begin{itemize}
\item We considered only one half spontaneous
breaking of global supersymmetries;
\item We used a very special parametrization of the coset, such that the super-space coordinates $\theta$'s do not
transform under spontaneously broken supersymmetry, while the physical fermions transform as the Goldstino
fields in the Volkov-Akulov model;
\item The superfield equations of motion in all cases are just the direct covariantization of the free ones.
\end{itemize}
Clearly, the component actions for other supersymmetry breaking patterns, as well as the actions for the super particles in 
another number of dimensions, can be similarly constructed starting from the corresponding super Poincar\'{e}
algebras.

It would be quite instructive to understand which new features will appear when we will replace the trivial, flat
target space by, for example, the AdS one. It seems, that the strategy will be the same and we are planning to report
the corresponding results elsewhere.

It is commonly understood that the superparticles and their actions are just
the simplest examples of the extended objects, from which only the
superstrings (and, probably, supermembranes) may be considered seriously as the theories possessing some physical applications. 
So, our main task was to analyze the components actions to find some common geometric structures, which can be further used in 
interesting models, including the Born-Infeld theories with extended supersymmetries.

The modern situation in this area may be regarded as a problematic one,
because the superspace approach meets many problems in the cases of extended supersymmetries, while the component approaches 
give such complicated actions, that any geometric intuition does not work.
Thus, our results just demonstrated that in the simplest theory with the
partial breaking of global supersymmetry, which is just the superparticle, there is a special choice of the components which 
dramatically simplifies the on-shell actions, still keeping the geometrically clear form of each terms in the action. 
We really believe that the known (and still unknown) actions
for super p-branes (and, hopefully, for super Born-Infeld theories) can
be re-formulated within the new set of variables with preserving the geometric meaning of the each object in the actions.

Furthermore, in the cases of superparticles the corresponding actions
have a unified structure.
Surprisingly, this structure is not sensitive to the number of
supersymmetries in the case of $D=3$.
Thus, the quantization of the superparticle in $D=3$ can be performed at once
for many systems with different numbers of supercharges.
Correspondingly, the spectrum of quantum states should have a common
structure too. We plan to analyze the quantum properties of the constructed actions elsewhere.

\section*{Acknowledgements}
We wish to acknowledge discussions with S.~Kuzenko and D.~Sorokin.

S.K. is grateful to the Laboratori Nazionali di Frascati for warm hospitality.
This work was partially supported by RFBR grants~11-02-01335-a, 13-02-91330-NNIO-a and 13-02-90602 Arm-a,
as well as by the ERC Advanced Grant no. 226455 \textit{``Supersymmetry, Quantum Gravity and Gauge Fields''}~(\textit{SUPER\-FIELDS}).

\end{document}